\documentclass[referee]{raa}            

\usepackage{graphicx,times}             
\usepackage{natbib}
\usepackage{amssymb,amsmath}
\bibpunct{(}{)}{;}{a}{}{,}
\usepackage[english]{babel}  
\usepackage{ctex} 

\usepackage[pagebackref=true]{hyperref}
\usepackage{color,soul}

\definecolor{delcolor}{rgb}{1,0,0}   
\definecolor{addcolor}{rgb}{0,0,1}   

\begin{document}

  \title{A Catalog of 1,408 Carbon-Enhanced Metal-Poor Stars from LAMOST DR11
}

   \volnopage{Vol.0 (202x) No.0, 000--000}      
   \setcounter{page}{1}          

   \author{Xianqi Liu
      \inst{1} 
   \and Xiangru Li
      \inst{1,*}\footnotetext{$*$Corresponding Author: Xiangru Li, lixiangru@scnu.edu.cn}
   \and Ziyu Fang
      \inst{1}
   } 

   \institute{South China Normal University, Guangdong Province, People's Republic of China; {\it 2023025175@m.scnu.edu.cn,xiangru.li@gmail.com, fangziyushiwo@126.com}\\
\vs\no
   }

\abstract{ Metal-poor (MP) stars are important targets for investigating the chemical evolution of the early universe. Among them, Carbon-Enhanced Metal-Poor (CEMP) stars have attracted extensive attention due to their rarity and astrophysical significance. Owing to their low occurrence rate, the identification of MP stars and CEMP stars remains a task of considerable scientific value. In this study, we investigate the search for CEMP stars based on the low-resolution stellar spectra from LAMOST (Large Sky Area Multi-Object Fiber Spectroscopic Telescope) DR11 and propose a deep-learning-based approach for this purpose. By analyzing the LAMOST DR11 spectral library, we identify 1,408 CEMP star candidates. For ease of reference and further use, we provide the estimated stellar parameters for these objects, including $T_\texttt{eff}$, $\log~g$, [Fe/H], and [C/H].
}

   \authorrunning{Liu et al.}            
   \titlerunning{A Catalog of 1,408 Carbon-Enhanced Metal-Poor Stars from LAMOST DR11 }  

   \maketitle

%
%
\section{Introduction}           
\label{sec:introduction}

Metal-poor (MP) stars are defined as stars whose outer atmospheres contain heavy elements (e.g., iron) at abundances far lower than those of the Sun. Since heavy elements are produced through nucleosynthesis processes inside stars, MP stars are generally older than the Sun and are believed to have formed in the early universe. Among MP stars, there exists a special subclass characterized by anomalously high carbon abundances, i.e., an enhancement of carbon relative to iron; these objects are referred to as Carbon-Enhanced Metal-Poor (CEMP) stars \citep{ARAA:Beers:2005,ApJ:Aoki:2007}. It has been found that the fraction of CEMP stars among MP stars increases with decreasing metallicity. At [Fe/H] $< -2.5$, the occurrence rate of CEMP stars can exceed 20\% \citep{placco2014carbon}. Owing to this high occurrence rate, CEMP stars are regarded as some of the most pristine objects in the universe. Consequently, the abundance patterns and origins of CEMP stars provide crucial clues for understanding early chemical evolution \citep{BonifacioAA2012}.

CEMP stars can be classified into several sub-classes, including CEMP-no, CEMP-s, CEMP-r, and CEMP-r/s stars, according to their distinct abundance patterns of neutron-capture elements \citep{ARAA:Beers:2005}. Nevertheless, the formation mechanisms of these sub-classes remain poorly understood and are still subject to considerable debate \citep{Fang_2025}. A comprehensive understanding of the properties and origins of these most ancient CEMP stars is crucial for probing the nature of the first generation of stars and the chemical enrichment history of the early Universe. However, systematic searches for CEMP stars continue to face substantial challenges, primarily due to the limited availability of suitable spectroscopic data and the intrinsic rarity of CEMP stars. Fortunately, with continuous advances in observational facilities and the implementation of large-scale sky surveys, the rapid release of massive spectroscopic datasets has created new opportunities for the systematic identification of CEMP stars. Representative survey projects include the Large Sky Area Multi-Object Fiber Spectroscopic Telescope (LAMOST) \citep{Zhao_2012,cui2012}, the Sloan Digital Sky Survey (SDSS) \citep{york2000sloan}, and the Global Astrometric Interferometer for Astrophysics (Gaia) \citep{brown2018gaia}. Among these extensive spectroscopic resources, low-resolution spectra are widely adopted for CEMP star searches. Compared with medium- and high-resolution spectroscopy, which requires significantly higher observational time and cost, low-resolution spectra provide much broader sky coverage and vastly larger sample sizes, thereby substantially increasing the probability of discovering rare CEMP stars. For instance, the LAMOST DR11 (v1.1) dataset contains more than 11.58 million low-resolution stellar spectra\footnote{\url{https://www.lamost.org/dr11/v1.1/doc/lr-data-production-description}}. Consequently, such a large-scale spectroscopic database offers an excellent opportunity for the systematic search and study of CEMP stars.

To efficiently identify CEMP stars from large-scale spectroscopic datasets, a variety of traditional approaches have been proposed, which can be broadly classified into two categories: line index methods based on spectral line detection and characterization, and template matching methods. The line index method extracts scalar features by integrating, computing weighted averages, or taking ratios of spectral flux within specific wavelength intervals, thereby compressing information related to the strength, width, or shape of spectral lines to characterize the physical properties of stellar objects. Since CEMP stars are typically characterized by low metallicity [Fe/H] and enhanced carbon abundance [C/Fe], this approach has been widely used to quantify spectral features associated with metal- and carbon-sensitive elements, particularly those exhibiting anomalously enhanced absorption in CEMP stars, such as the CH G-band, CN bands, and C2 bands. These distinctive features play an important role in distinguishing CEMP stars from normal stellar populations. For example, the HK survey in the OBJECTIVE-PRISM SURVEY \citep{1985AJBeers,1992AJBeers} and the Hamburg/ESO Survey (HES) \citep{2006ApJFrebel,2008AAChristlieb} searched for target objects based on weak Ca II absorption lines; \citet{1999AJBeers,2008AAChristlieb,2010AJPlacco,Placco2011AJ,LiHaiNing2018ApJS} identified CEMP stars using the CH G-band; and \citet{2018MNRASCotar} employed the Swan bands for candidate selection. Overall, the line index method offers strong physical interpretability; however, it relies heavily on the selection of spectral lines and wavelength ranges and typically requires relatively high spectral resolution, which limits its applicability in large-scale, low-resolution spectroscopic surveys.

Template matching methods infer the physical parameters or classifications of astronomical objects by comparing observed spectra with pre-constructed template spectra and identifying the best-matching template through $\chi^2$ minimization or other similarity metrics. This approach has been widely applied in the search for CEMP stars. For example, \citet{2013LeeAJ} matched Sloan Digital Sky Survey (SDSS) and Sloan Extension for Galactic Understanding and Exploration (SEGUE) spectra with synthetic spectral templates to estimate [C/Fe] within wavelength regions containing the CH G-band feature; \citet{ApJ:Lihaining:2015} identified CEMP candidates in the LAMOST DR1 dataset by directly comparing normalized observed spectra with synthetic templates; \citet{2017AAAguado} proposed a two-step selection strategy in which candidate stars are first screened using the CaII resonance lines to estimate [Fe/H], followed by a template matching procedure applied to local wavelength regions around the Balmer lines to remove spurious candidates; and \citet{LiHaiNing2018ApJS} compared LAMOST DR3 spectra with synthetic templates generated using the SPECTRUM synthesis code \citep{gray1994calibration} to search for extremely metal-poor stars and CEMP stars. Overall, the performance of template matching methods strongly depends on the completeness and quality of the template library; insufficient template coverage or significant discrepancies between the templates and the observed data may reduce the accuracy and reliability of the search results.

With the rapid development of artificial intelligence techniques, an increasing number of studies have attempted to employ machine learning methods for the search of CEMP stars. Stellar spectra not only contain sequential information in the wavelength domain, but also encode rich frequency-domain structures, which can be extracted through time–frequency analysis techniques such as the Short-Time Fourier Transform (STFT) \citep{esmaeilpour2022multidiscriminator}. These frequency-domain features are crucial for characterizing stellar chemical abundances and physical states. However, most existing machine learning approaches rely primarily on a single data view, namely the raw spectral sequence, and fail to jointly exploit and integrate the time–frequency information embedded in stellar spectra. This limitation restricts both the accuracy of stellar parameter estimation and the effectiveness of CEMP star identification. For example, some studies \citep{2017ApJSCarbon, 2018MNRASCotar} performed clustering or classification based solely on the original spectral sequences, thereby neglecting their frequency-domain structures. Other supervised learning-based methods \citep{2023MNRASLucey, Xie_2021, ardern2025predicting} were able to identify candidate CEMP stars, yet exhibited limitations in the completeness or accuracy of parameter predictions, and still did not effectively incorporate the time–frequency representations of stellar spectra.  More critically, low-resolution spectra from large-scale surveys are generally accompanied by low signal-to-noise ratios, making it particularly challenging to accurately extract faint spectral features and frequency-domain structures from single-view spectra. Moreover, frequency-domain features, which are correlated with subtle variations in chemical abundances, remain largely underutilized in existing machine learning frameworks.

To address the aforementioned limitations, we propose a dual-view weak-signal learning framework tailored for low signal-to-noise ratio (SNR) spectra. The methodological novelty lies in the systematic integration of two complementary representations: the original spectral sequence in the wavelength domain and its time–frequency representation in the frequency domain. This design alleviates the reliance of conventional approaches on a single wavelength-domain representation at the modeling level. From both physical and signal-processing perspectives, key chemical signatures in stellar spectra (e.g., the CH G-band feature of carbon-enhanced metal-poor stars) are manifested not only as localized absorption line profiles in the wavelength domain, but also as characteristic frequency components and structural distribution patterns in the spectral signal. Under low-SNR conditions, weak absorption features in the wavelength domain may be partially obscured by noise or suffer morphological degradation, thereby limiting the stability of single-view modeling approaches. By constructing a time–frequency representation via the Short-Time Fourier Transform (STFT), the local variations and global modulation patterns of spectral signals can be re-characterized from the perspective of frequency structures, providing complementary feature descriptions beyond those available in the wavelength domain. Based on this motivation, the proposed framework introduces an Attention Calibration Recurrent module (ACR) to model local spectral-line structures and their sequential dependencies in the wavelength domain, together with a Parallel Multi-scale Time–Frequency State Space module (PMTF) to extract multi-scale structural information in the frequency domain. The two branches are jointly optimized and fused within a unified framework, enabling cross-view feature complementarity and weak-signal enhancement. To examine the relative contribution of each branch in the dual-view design, we conduct systematic ablation experiments (see Section ~\ref{sec:parameterEstimation_evaluation}). The results indicate that the dual-view architecture consistently outperforms either single-view variant in key physical parameter estimation tasks. This suggests that the frequency-domain representation provides non-redundant and practically contributive information, improving model stability and robustness under low-SNR conditions. This multi-view modeling strategy offers a robust approach for identifying CEMP stars in large-scale, low-resolution spectroscopic surveys.

The remainder of this paper is organized as follows. Section~\ref{sec:Workflow_Data_Preprocessing} introduces the overall workflow of the CEMP star search, including the datasets used and the data preprocessing procedures. Section~\ref{sec:parameterEstimation} presents the proposed method for estimating stellar spectral parameters, and these parameter estimation results are employed to screen and identify CEMP star candidates. Section~\ref{sec:CEMPSearch} reports the search results of CEMP star candidates. Section~\ref{sec:ParameterEvaluation} compares the consistency of the parameter estimation results obtained in this work with those from GALAH. Finally, Section~\ref{sec:conclusion} provides a brief summary of the proposed CEMP star search scheme and its corresponding results.

\section{Methodology, Data, and Preprocessing}
\label{sec:Workflow_Data_Preprocessing}

This section briefly outlines the overall workflow for CEMP star searching and describes the reference dataset and preprocessing methods used.

\subsection{Overall Process}
\label{sec:Workflow}

This study aims to search for CEMP star candidates from the LAMOST DR11 low-resolution spectral database. For a given low-resolution stellar spectrum, we first estimate its stellar parameters $T_\texttt{eff}$, $\log~g$, [Fe/H], and [C/H]; then select MP star candidates based on the criterion [Fe/H] $< -1$; finally, the retrieved MP star candidates are classified into CEMP star candidates and Carbon-Normal Metal-Poor star candidates according to the criteria in \citet{ApJ:Aoki:2007}:
\begin{equation}\label{eq:CEMP_define}
\left\{
\begin{array}{lc}
\log \left(L / L_{\odot}\right) \leq 2.3 \land {[C / F e] \geq+0.7} \\
\log \left(L / L_{\odot}\right)>2.3 \land {[C / F e] \geq+3.0-\log \left(L / L_{\odot}\right)}
\end{array}
\right.
\end{equation}
where
\begin{align*} L/L_\odot & \propto (R/R_\odot)^2(T_\texttt{eff}/T_{\texttt{eff}\odot})^4\\
                     & \propto (M/M_\odot) (g/g_\odot)^{-1} (T_\texttt{eff}/T_{\texttt{eff}\odot})^4
\end{align*}
$M_\odot$, $g_\odot$ and $T_{\texttt{eff}\odot}$ represent the mass, surface gravity, and effective temperature of the sun, respectively.

It should be noted that the adoption of [Fe/H] $< -1.0$ as the initial screening threshold for metal-poor (MP) stars in the above workflow is motivated by the following considerations. First, the measurement of [Fe/H] from LAMOST low-resolution spectra is subject to non-negligible random uncertainties, particularly for spectra with low signal-to-noise ratios. If a more stringent threshold of [Fe/H] $< -2.0$ were applied at the initial stage, genuine extremely metal-poor targets with intrinsic metallicities below $-2.0$ but slightly overestimated measured values could be prematurely excluded. Second, the primary objective of this study is to construct a CEMP candidate catalog with high completeness. A relatively relaxed pre-selection threshold helps retain as many potential targets as possible at the early stage, thereby reducing the risk of missing bona fide CEMP stars. Finally, the pre-selected sample subsequently undergoes rigorous CEMP classification criteria as well as additional filtering procedures, including spatial distribution and kinematic selections (see Section~\ref{sec:CEMPSearch_results} for details). These subsequent steps effectively remove possible metal-rich contaminants introduced by the relaxed initial threshold. Our empirical analysis demonstrates that, after applying the full pipeline, cross-validation with high-resolution spectroscopic data from APOGEE DR17 reveals no metal-rich contaminants with [Fe/H] $\geq -1.0$ in the final catalog. In other words, the false positive rate (i.e., the fraction of metal-rich stars misclassified as metal-poor stars) is 0\% (see Section~\ref{sec:CEMPSearch_results}). This result confirms that the adopted threshold ensures high sample completeness without introducing significant contamination.

\subsection{Reference Dataset for CEMP Star Search}
\label{sec:Reference_set}

The reference dataset for stellar parameter estimation used in this study is adopted directly from the work of \citet{liu2025weaksignallearningdataset}. This dataset is based on LAMOST DR11 v1.1 low-resolution spectra, with precise stellar atmospheric parameters ($T\text{eff}$, $\log~g$, [Fe/H], [C/H]) obtained by cross-matching several high-fidelity catalogs (APOGEE DR17, LAMOST-Subaru, SAGA, and a VMP catalog). To optimize the search performance for metal-poor and CEMP stars, the original work performed downsampling to balance the class distribution. The final dataset comprises 13,158 spectra spanning a broad parameter space (e.g., $\text{[Fe/H]} \in [-4.37, 0.50]~\text{dex}$). It has been partitioned into training ($S^{stars}_{tr}$), validation ($S^{stars}_{va}$), and test ($S^{stars}_{te}$) sets in a 7:1:2 ratio, while preserving class distribution consistency. The detailed construction process, quality control, and statistical analysis of the dataset are described in \citet{liu2025weaksignallearningdataset}.

Although the scale of the reference dataset ($N=13,158$) is relatively small compared to the search pool ($N \approx 9.7 \times 10^6$), the training set exhibits sufficient representativeness across key parameter regimes to support reliable generalization of the model. As shown in Table~\ref{tab:training_distribution}, the training set demonstrates continuous and adequate coverage in the low-metallicity region: the Extremely Metal-Poor (EMP) region with $\text{[Fe/H]} < -3.0$ contains 158 samples (1.72\%), while the Very Metal-Poor (VMP) with $\text{[Fe/H]} < -2.0$ comprises 2,276 samples (24.72\%). This indicates that the model can learn relevant spectral features from a considerable number of low-metallicity samples. At the high-temperature end, the training set also spans a broad range: samples with $T_{\text{eff}} \geq 5500$ K account for 28.81\%, among which 1,149 samples (12.48\%) have $T_{\text{eff}} \geq 6000$ K. Such a distribution ensures that the model is trained not only on common samples with higher metallicity or lower temperature, but across the entire parameter space, including both low-metallicity and high-temperature regimes. Therefore, our model possesses the capability to perform reasonable parameter estimation for unknown spectra, rather than merely extrapolating from the higher-metallicity portion of the training samples.

\begin{table}[htbp]
    \begin{center}
    \caption{Sample distribution of the training set ($S^{stars}_{tr}$) across key parameter regimes}
    \label{tab:training_distribution}
    
    \begin{tabular}{ccc}
    \hline\noalign{\smallskip}
    Parameter Regime & Count & Percentage \\
    \hline\noalign{\smallskip}
    
    $[\mathrm{Fe/H}] < -3.0$ & 158 & 1.72\% \\
    $-3.0 \leq [\mathrm{Fe/H}] < -2.0$ & 2118 & 23.00\% \\
    $-2.0 \leq [\mathrm{Fe/H}] < -1.0$ & 2531 & 27.48\% \\
    $-1.0 \leq [\mathrm{Fe/H}] < 0.0$ & 3033 & 32.93\% \\
    $[\mathrm{Fe/H}] \geq 0.0$ & 1370 & 14.88\% \\
    
    \noalign{\smallskip}\hline\noalign{\smallskip}
    
    $T_{\mathrm{eff}} < 4000$\,K & 43 & 0.47\% \\
    $4000\,\mathrm{K} \leq T_{\mathrm{eff}} < 4500\,\mathrm{K}$ & 1081 & 11.74\% \\
    $4500\,\mathrm{K} \leq T_{\mathrm{eff}} < 5000\,\mathrm{K}$ & 3344 & 36.31\% \\
    $5000\,\mathrm{K} \leq T_{\mathrm{eff}} < 5500\,\mathrm{K}$ & 2089 & 22.68\% \\
    $5500\,\mathrm{K} \leq T_{\mathrm{eff}} < 6000\,\mathrm{K}$ & 1504 & 16.33\% \\
    $T_{\mathrm{eff}} \geq 6000$\,K & 1149 & 12.48\% \\
    
    \noalign{\smallskip}\hline
    \end{tabular}
    \end{center}
\end{table}

\subsection{Data Preprocessing}
\label{sec:data_process}

This paper follows the same data preprocessing pipeline as \citet{liu2025weaksignallearningdataset} for the LAMOST DR11 v1.1 low-resolution spectra to mitigate biases introduced by factors such as scattering, reflection, and instrumental anomalies, thereby improving the reliability of subsequent parameter estimation and star selection. The main steps include: wavelength correction to the rest frame using the radial velocity provided by LAMOST; linear interpolation resampling in logarithmic space within the largest common wavelength range to unify the spectral format; application of median filtering for noise reduction; continuum normalization via iterative fitting of a high-order polynomial; and finally, secondary denoising  and spectrum-wise standardization. 

\begin{figure}[ht]
    \centering
    \subfloat[CEMP star spectrum ({[}C/Fe{]}=0.76)]{%
        \includegraphics[width=0.48\textwidth]{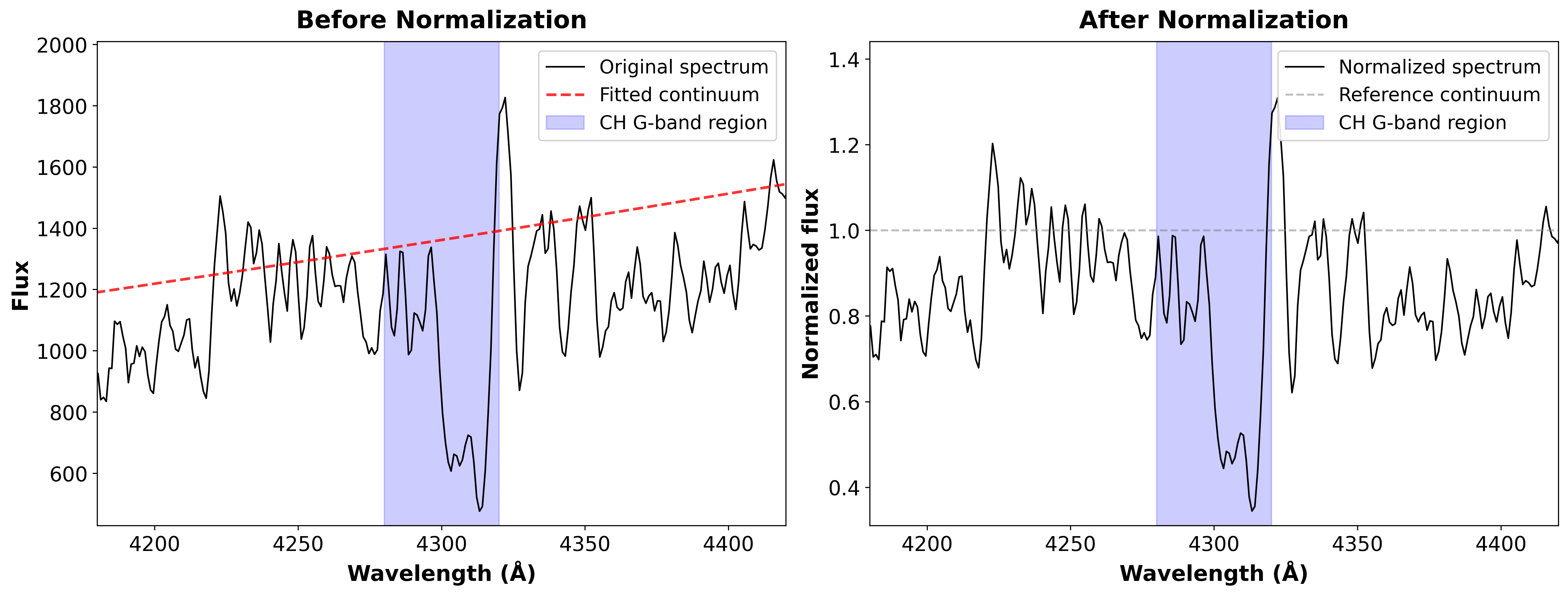}%
        \label{fig:normalization_verification_327303219}%
    }
    \subfloat[CEMP star spectrum ({[}C/Fe{]}=0.98)]{%
        \includegraphics[width=0.48\textwidth]{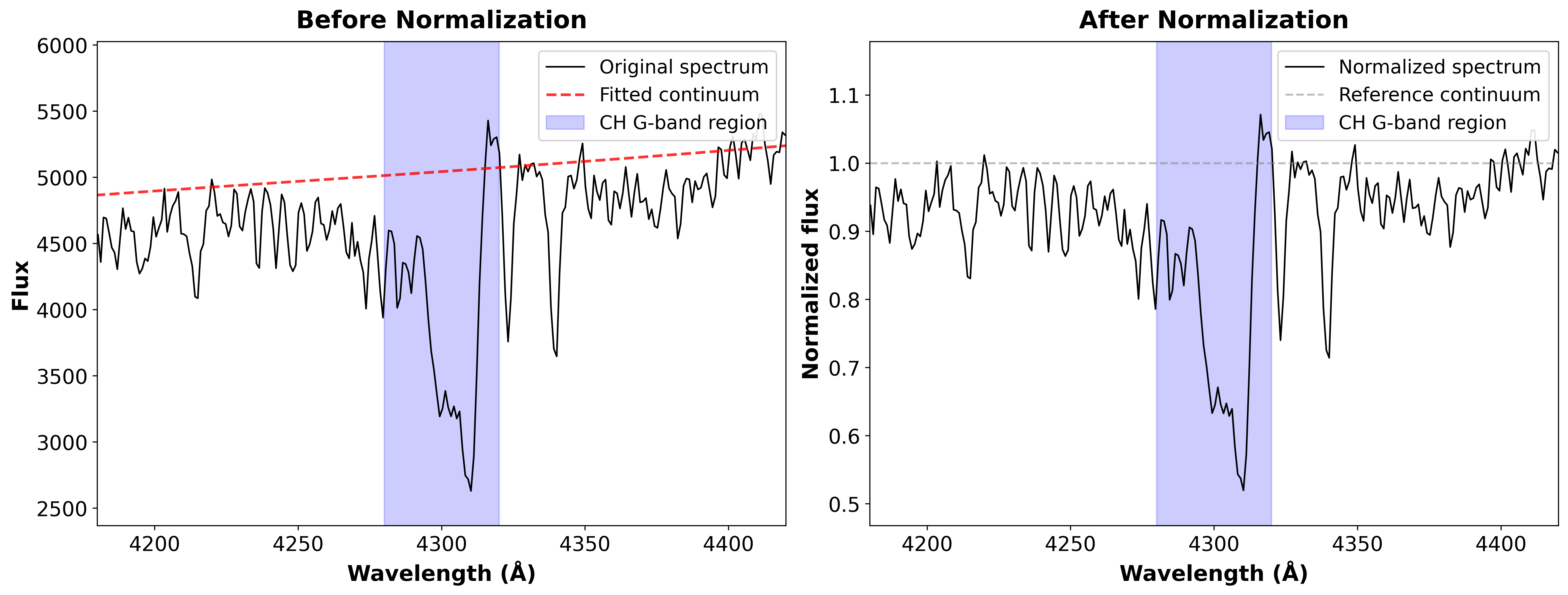}%
        \label{fig:normalization_verification_208207221}%
    } \\
    \subfloat[CEMP star spectrum ({[}C/Fe{]}=1.52) ]{%
        \includegraphics[width=0.48\textwidth]{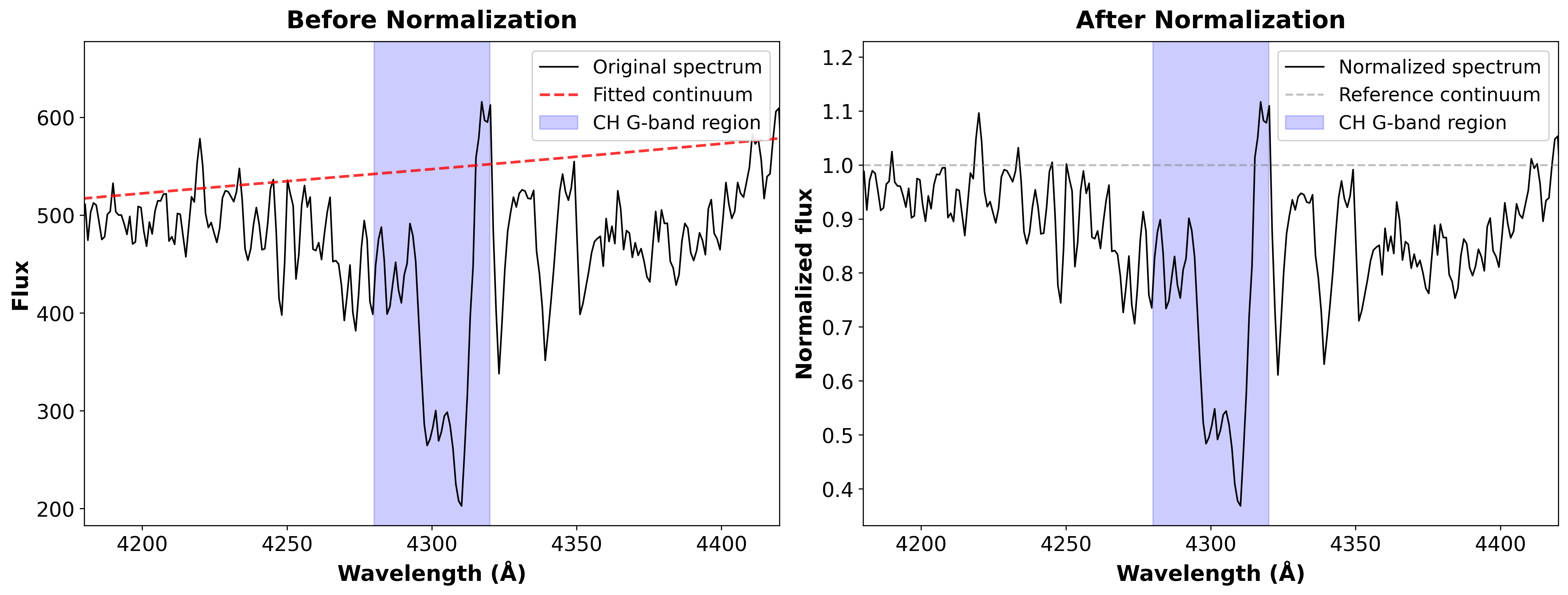}%
        \label{fig:normalization_verification_66610035}%
    }
    \subfloat[CEMP star spectrum ({[}C/Fe{]}=2.98) ]{%
        \includegraphics[width=0.48\textwidth]{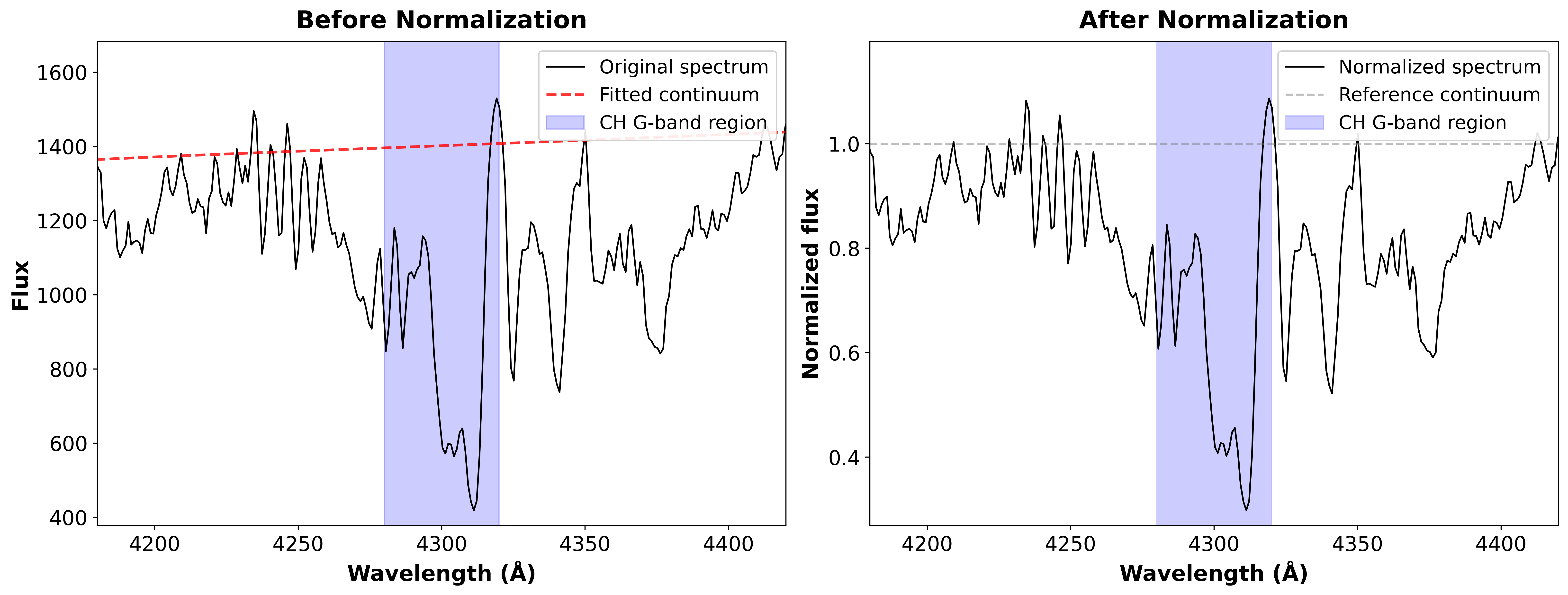}%
        \label{fig:normalization_verification_120805189}%
    }
    \caption{Comparison of four CEMP stellar spectra before and after continuum normalization, illustrating the preservation of the CH G-band (approximately 4280--4320\,\AA, shaded in purple) across a range of [C/Fe] values. The black curve shows the original flux, the red dashed line the fitted continuum, while the light gray horizontal dashed line indicates the reference continuum after normalization ($y=1.0$).}
    \label{fig:normalization_verification}
\end{figure}

In low-resolution spectra of metal-poor stars—particularly CEMP stars—the key carbon-sensitive features, such as the CH G-band located at approximately 4300\,\AA, can be extremely weak. As a critical step in data preprocessing, continuum normalization carries a potential risk: these subtle absorption features may be mistakenly treated as local fluctuations of the continuum and consequently suppressed or artificially flattened. To verify that our preprocessing pipeline reliably preserves such features across the entire dataset, we selected four representative CEMP spectra based on their [C/Fe]. The chosen stars cover a wide range of [C/Fe] values, from just above the CEMP threshold to the highest values present in our dataset. This selection ensures that the normalization procedure is tested under varying strengths of the carbon feature, thereby providing a comprehensive assessment of its robustness. The [C/Fe] values of the selected stars are indicated in the subcaptions of Figure~\ref{fig:normalization_verification}.

Figure~\ref{fig:normalization_verification} presents the spectral comparison before and after normalization for these four stars. After normalization, the absorption depression in the CH G-band region (highlighted by the purple shaded area) remains clearly identifiable in all cases, with its morphology and depth consistent with those in the original spectra. This result demonstrates that the iterative high-order polynomial fitting method adopted in this work is capable of performing effective continuum normalization while faithfully retaining the critical carbon-sensitive spectral information. The consistent preservation observed across these four representative stars—covering a wide range of [C/Fe] values—indicates that the method reliably retains weak carbon features throughout the entire dataset. Consequently, it provides a reliable data foundation for subsequent morphology-based machine learning analysis of stellar spectra. Detailed preprocessing procedures can be found in \citet{liu2025weaksignallearningdataset}.

\section{Stellar Spectroscopic Parameter Estimation}
\label{sec:parameterEstimation}

This chapter estimates stellar atmospheric parameters ($T_{\texttt{eff}}$, $\log~g$, [Fe/H], [C/H]) and performs stellar classification based on the Parallel Dual-View Fusion Network architecture from \citet{liu2025weaksignallearningdataset}, by extending its output branches and employing ensemble learning.

\subsection{Parallel Dual-View Fusion Network (PDVFN)}
\label{sec:parameterEstimation_model}

The ultimate goal of this work is to search for CEMP star candidates from LAMOST low-resolution spectra. The core of this endeavor is to construct a high-precision stellar parameter estimation model to derive key parameters such as $T_\text{eff}$, $\log~g$, [Fe/H], and [C/H]. To this end, we adopt the Parallel Dual-View Fusion Network (PDVFN) architecture proposed by \citet{liu2025weaksignallearningdataset}, a model that jointly learns from the spectral sequence and its frequency-domain representation. To enhance the model's sensitivity to weak features associated with CEMP stars (e.g., the CH G-band), we introduce a key modification: alongside the original single parameter estimation branch, we add a parallel auxiliary classification branch. Both branches share the front-end feature extraction layers. The output of this classification branch is not directly used for the final CEMP star identification. Its primary role is to guide the shared feature extraction layers during training—through its loss function—to focus more on discriminative spectral features that can distinguish CEMP stars from non-CEMP stars (e.g., weak absorption lines related to carbon abundance). This design enables the primary parameter estimation branch to perform regression based on richer and more discriminative features, thereby improving the estimation accuracy of key parameters like $T_\text{eff}$, $\log~g$, [Fe/H], and [C/H].

When training such a network with an auxiliary branch, balancing the loss functions of its components is crucial to prevent the auxiliary task from interfering with the primary task (i.e., ``negative transfer''). For the primary parameter estimation branch, we employ the Negative Log-Likelihood (NLL) loss, denoted as $\mathit{L_{param}}$. This loss function simultaneously optimizes both the mean and variance of the predictions, thereby enhancing the model's robustness under high-noise conditions. For the auxiliary classification branch, to address the class imbalance issue arising from the relatively small number of CEMP star samples, we adopt the Focal Loss, denoted as $\mathit{L_{focal}}$. This loss reduces the contribution weight of easy-to-classify samples to the total loss, forcing the model to focus more on hard-to-distinguish samples. The final training objective of the model is composed of these two loss components:
\begin{equation}
    \mathit{loss} = \mathit{L}_{param} + \mathit{L}_{focal}
\end{equation}

It is important to clarify that in our model design, the shared feature extraction layer is the core component, and its primary beneficiary is the main parameter estimation task. The loss from the auxiliary classification branch, $\mathit{L_{focal}}$, directly influences the parameter updates of the shared layers through gradient back-propagation, driving the network to learn feature patterns useful for classification (i.e., distinguishing CEMP stars). These enhanced features (e.g., representations more sensitive to the CH G-band) are in turn utilized by the parameter estimation branch, thereby directly improving the estimation accuracy of key parameters such as [C/H]. To quantitatively verify the benefit of this design, we conducted an ablation experiment comparing the performance of the full dual-branch model against a model retaining only the parameter regression branch (i.e., removing the auxiliary classification branch and its loss). The results are shown in Table~\ref{tab:ablation_mae}. It can be seen that the introduction of the auxiliary classification branch leads to improved estimation accuracy across all four parameters ($T_\text{eff}$, $\log~g$, [Fe/H], and [C/H]). These results clearly demonstrate that the auxiliary branch, by guiding the shared layers to learn more discriminative features, effectively enhances the performance of the main parameter estimation task, confirming that the dual-branch design is superior to the single parameter regression branch. Apart from the aforementioned addition of the auxiliary branch and its associated loss functions, the core network architecture, feature fusion mechanism, and main training strategy of the model remain consistent with those described in \citet{liu2025weaksignallearningdataset}.

\begin{table}[htbp]
    \centering
    \caption{Comparison of MAE between the dual-branch model and the regression-only model}
    \label{tab:ablation_mae}
    
    \begin{tabular}{lcccc}
    \hline\noalign{\smallskip}
    Model & $T_{\mathrm{eff}}$ & $\log~g$ & $[\mathrm{Fe/H}]$ & $[\mathrm{C/H}]$ \\
    \noalign{\smallskip}\hline\noalign{\smallskip}
    
    Regression-only  & 103.7538 & 0.2538 & 0.1515 & 0.2044 \\    
    Dual-branch      & \textbf{100.7120} & \textbf{0.2509} & \textbf{0.1486} & \textbf{0.1974} \\
    
    \noalign{\smallskip}\hline
    \end{tabular}
    
    \smallskip
    \parbox{0.6\columnwidth}{ 
    \footnotesize
    \textit{Note.} ``Regression-only'' refers to the model with only the parameter regression branch (i.e., without the auxiliary classification branch); ``Dual-branch'' refers to the full model with both parameter regression and auxiliary classification branches. MAE values are shown for each parameter; lower values indicate better performance, and the optimal results are highlighted in bold.
    }
\end{table}

\subsection{Ensemble Strategy}
\label{sec:parameterEstimation_ensemble}

To enhance the robustness against complex noise in low-resolution spectra and improve the generalization capability of parameter estimation, this work employs an ensemble learning strategy. We adopt the Blending method \citep{koren2009bellkor}, a simplified version of Stacking \citep{wolpert1992stacked}, to obtain more stable estimates by combining predictions from multiple Parallel Dual-View Fusion Networks (PDVFN). Specifically, multiple networks are first trained on the training set $S^{stars}_{tr}$ as primary learners. Then, their predictions on the validation set $S^{stars}_{va}$ are used to construct a secondary dataset, on which a multivariate linear regression model is trained as the secondary learner to produce the final prediction. This approach effectively integrates complementary information from multiple models, enhancing both stability and accuracy of the system.

\subsection{Model Evaluation}
\label{sec:parameterEstimation_evaluation}

\begin{figure}[ht]
    \centering
    \includegraphics[width=\textwidth]{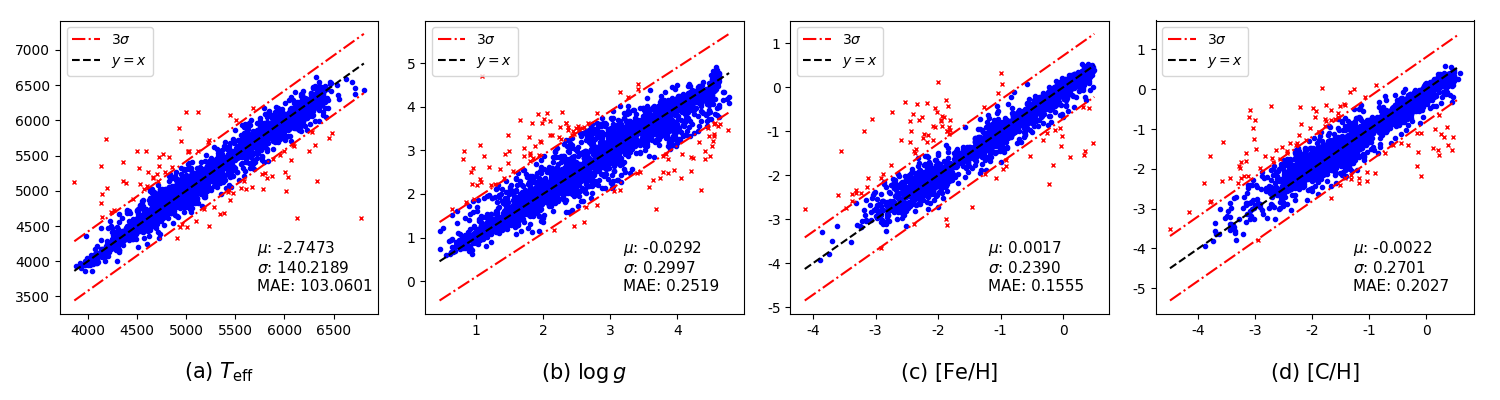}
    \caption{The performance of parameter estimation. The horizontal axis represents the reference values and the vertical axis represents the estimated values. These results are computed from the test set. The black dashed lines represent the reference line for theoretical consistency, while the red dashed lines represent the 3$\sigma$ reference lines. The $\mu$, $\sigma$ are the mean and standard deviation of the difference between the parameter estimation and its corresponing reference on test spectra. MAE is the mean absolute error/difference between the parameter estimation and its corresponing reference on test spectra.}
    \label{fig:parameterEstimationResults}
\end{figure}

To assess the discrepancies between the model predictions and the true labels, we conducted a detailed comparison of the estimated values with the ground truth in the test set, as shown in Figure~\ref{fig:parameterEstimationResults}. For $T_\texttt{eff}$, $\log~g$, [Fe/H], and [C/H], the proportions of outliers beyond the 3$\sigma$ range are only 3.15\%, 4.03\%, 3.12\%, and 3.61\%, respectively. This indicates a high consistency between the model predictions and the true values. It is worth noting that some spectra in the LAMOST dataset are of relatively low quality, resulting in certain estimation errors, which are the primary cause of these outliers. To mitigate this instability, a high-confidence stellar catalog has been compiled in Section~\ref{sec:CEMPSearch_results} to facilitate more precise follow-up studies of CEMP stars. Further relevant evaluations can be found in Section~\ref{sec:ParameterEvaluation}.

To quantitatively assess the contribution of each component in the dual-view framework, we conducted ablation experiments by removing either the vector-view branch (ACR) or the time–frequency-view branch (PMTF). The experimental results (Table~\ref{tab:ablation_results}) clearly demonstrate that the complete dual-view model (ACR+PMTF) achieves the lowest mean absolute error (MAE) for all astrophysical parameters ($T_{\text{eff}}$, $\log~g$, [Fe/H], and [C/H]). Notably, compared with the baseline model that relies solely on time-domain features (ACR only), the incorporation of frequency-domain information (PMTF) leads to consistent and substantial performance improvements. For instance, the MAE of [Fe/H] is significantly reduced from 0.1865 dex to 0.1487 dex (a relative improvement of 20.3\%), while the MAE of [C/H] decreases from 0.2479 dex to 0.2007 dex (a relative improvement of 19.0\%).  These quantitative results demonstrate the effectiveness of incorporating frequency-domain information for improving parameter estimation accuracy.

\begin{table}
    \centering
    \caption{Ablation experiment results}
    \label{tab:ablation_results}
    
    \begin{tabular}{ccrrrr}
    \hline\noalign{\smallskip}
    \multicolumn{2}{c}{Ablation components} &
    \multicolumn{4}{c}{Parameter estimation} \\
    \noalign{\smallskip}\hline\noalign{\smallskip}
    
    ACR & PMTF &
    MAE\_${T_{\mathrm{eff}}}$ &
    MAE\_${\log~g}$ &
    MAE\_${[\mathrm{Fe/H}]}$ &
    MAE\_${[\mathrm{CH}]}$ \\
    
    \noalign{\smallskip}\hline\noalign{\smallskip}
    
    $\times$ & $\checkmark$ & 158.3333 & 0.4026 & 0.3160 & 0.3260 \\
    $\checkmark$ & $\times$ & 118.4524 & 0.2854 & 0.1865 & 0.2479 \\
    $\checkmark$ & $\checkmark$ &
    \textbf{100.7120} & \textbf{0.2509} & \textbf{0.1486} & \textbf{0.1974}\\    
    
    \noalign{\smallskip}\hline
    \end{tabular}
    
    \smallskip
    \parbox{0.8\columnwidth}{
    \footnotesize
    \textit{Note.} ``ACR'' denotes the vector-view branch, which processes the original 1D spectral vectors and represents temporal feature modeling; ``PMTF'' denotes the time-frequency view branch, which processes the time-frequency views generated by Short-Time Fourier Transform (STFT) and represents frequency-domain feature modeling. ``$\checkmark$'' indicates that the branch is enabled, and ``$\times$'' indicates that it is removed. MAE\_${T_{\mathrm{eff}}}$ denotes the mean absolute error between the model's estimated value and the label for the effective temperature $T{\mathrm{eff}}$. MAE\_${\log~g}$, MAE\_${[\mathrm{Fe/H}]}$, and MAE\_${[\mathrm{CH}]}$ represent the corresponding error metrics for surface gravity $\log~g$, metallicity $[\mathrm{Fe/H}]$, and carbon abundance $[\mathrm{CH}]$, respectively. Lower MAE values indicate better performance, and the optimal results are highlighted in bold.
    }
\end{table}

\section{CEMP Star Search}
\label{sec:CEMPSearch}

\subsection{Search Results for CEMP Star Candidates}
\label{sec:CEMPSearch_results}

Following the CEMP star search workflow described in Section~\ref{sec:Workflow}, after completing the estimation of the stellar parameters $T_{\texttt{eff}}$, $\log~g$, [Fe/H], and [C/H] (Section~\ref{sec:parameterEstimation}), we first select metal-poor star candidates according to the criterion [Fe/H] $\leq -1$. Subsequently, based on the discriminant defined in Eq.~\eqref{eq:CEMP_define}, these candidates are further classified into CEMP candidates and Carbon-Normal Metal Poor candidates.

\begin{figure}[ht]
    \centering

    \subfloat[Distribution of CEMP Candidates]{%
        \includegraphics[width=0.46\textwidth]{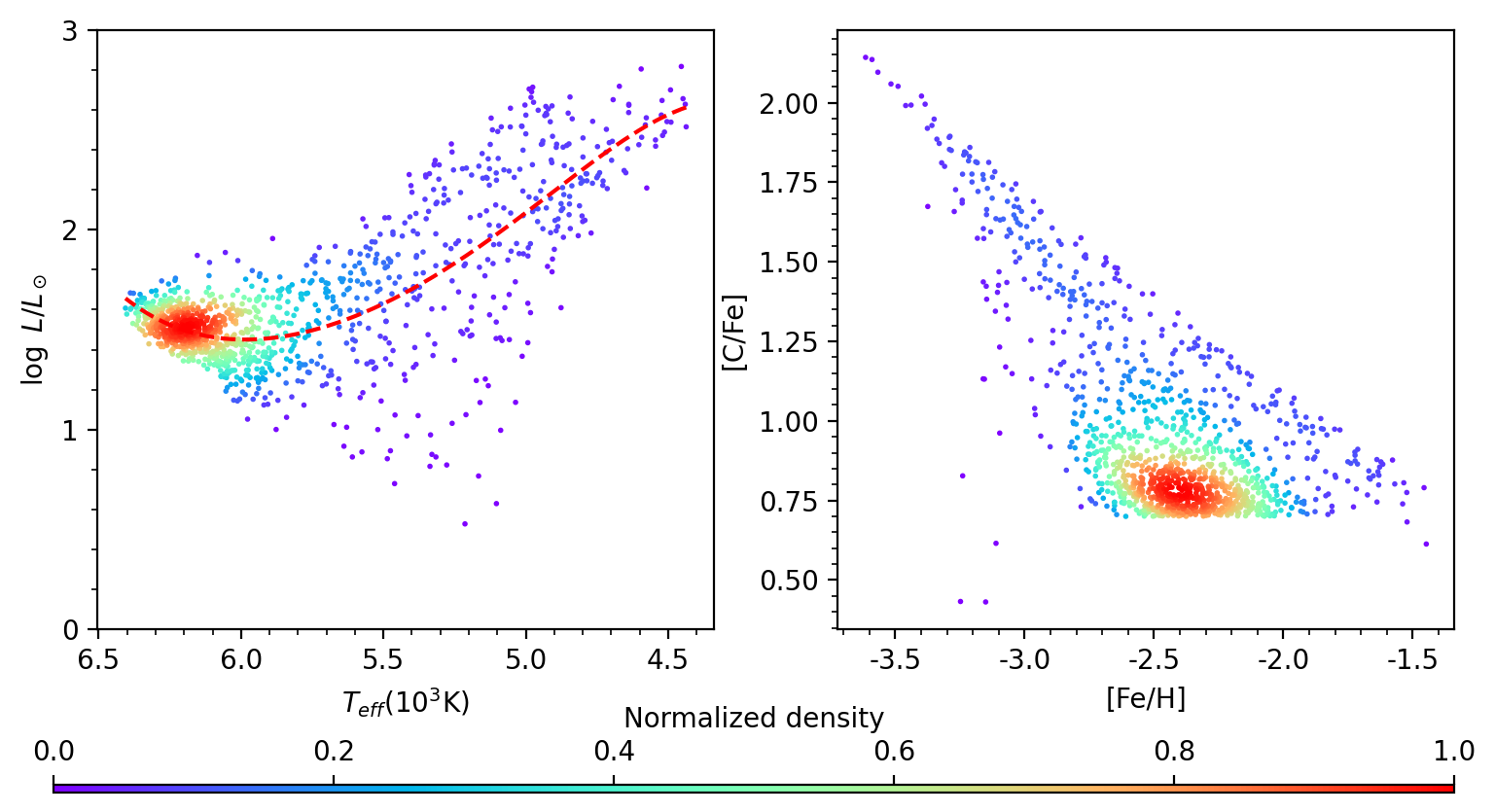}%
        \label{fig:CEMP_distribution}%
    }\hspace{0.01cm}%
    \subfloat[Occurrence Rate of CEMP Candidates]{%
        \includegraphics[width=0.46\textwidth]{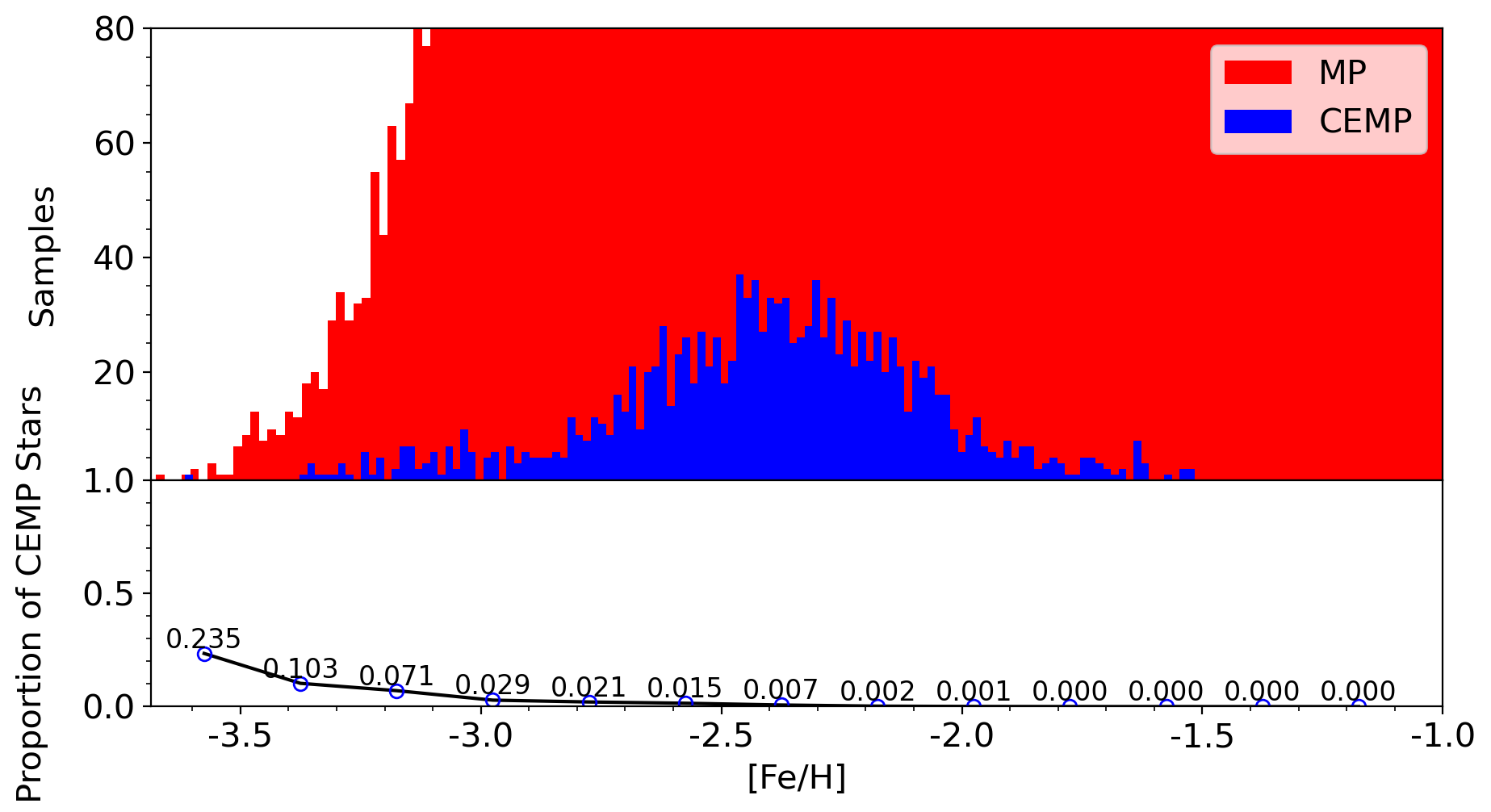}%
        \label{fig:CEMP_proportion}%
    }
    \caption{(a) Distribution of 1,408 CEMP candidates in the parameter space; (b) Occurrence rate of the CEMP candidates, where bins with sample sizes exceeding 80 are not shown. [C/Fe] = [C/H] − [Fe/H].}
    \label{fig:CEMP_distribution_occurrence_rate}
\end{figure}

After conducting a systematic search of 9,772,245 low-resolution spectra with relatively high signal-to-noise ratios (SNR$_g > 5$) from LAMOST DR11, we further refined the candidate sample by incorporating astrometric and kinematic information from Gaia DR3 \citep{gaia2023gaia,prusti2016gaia}. First, based on the spatial distributions of the candidates (Galactic latitude and vertical distance from the Galactic plane), we selected 5,315 objects exhibiting spatial characteristics consistent with those of halo stars. Subsequently, a further screening based on kinematic properties (proper motions and space velocities) was performed,  yielding an initial sample of 1,431 CEMP candidates whose spatial distributions and kinematic behaviors are both consistent with halo star populations. To further enhance sample purity, we cross-checked these candidates using the official metallicity values provided by LAMOST DR11, identifying 23 potential contaminants with metallicity $\geq -1.0$, which were then removed. This refinement resulted in a final sample of 1,408 high-confidence CEMP candidates. These candidates are predominantly concentrated in the extremely low-metallicity regime: Very Metal-Poor (VMP) stars with [Fe/H] $< -2$ constitute the majority, totaling 1,302 objects, while 104 candidates are classified as Extremely Metal-Poor (EMP) stars with [Fe/H] $< -3$. The distributions of these candidates in the relevant parameter space are illustrated in Figure~\ref{fig:CEMP_distribution}. To further assess the purity of our catalog, we cross-matched the final sample with high-resolution spectroscopic data from APOGEE DR17. None of the matched sources exhibit [Fe/H] $\geq -1.0$, yielding a false-positive rate of 0\%. This provides independent and robust external validation of the high purity achieved by our selection strategy. 

It should be noted that, according to the official documentation from LAMOST\footnote{\url{https://www.lamost.org/dr11/v1.1/doc/lr-data-production-description}}, not all low-resolution spectra are suitable for high-precision inversion of stellar atmospheric parameters. Considering that this limitation may affect the classification of CEMP stars, we assign confidence levels to the corresponding CEMP star candidates. Specifically, candidates whose spectra have a signal-to-noise ratio (SNR) below 35 and a LAMOST redshift value ``Z'' of -9999 (indicating poor spectral quality) are classified as low-confidence, while the remaining 444 candidates are regarded as high-confidence.

The HR diagram in the left panel of Figure~\ref{fig:CEMP_distribution} reveals the evolutionary track of CEMP stars, transitioning from the main sequence through the subgiant phase and ultimately to the red giant stage. Although Eq.~\eqref{eq:CEMP_define} does not explicitly constrain the relationship between [C/Fe] and [Fe/H], the search results shown in the right panel of Figure~\ref{fig:CEMP_distribution} indicate that the correction for stars with [C/Fe] $< 0.7$ is only partially effective, which is consistent with the distribution observed in the reference dataset. This phenomenon likely arises because some stars, during their giant phase, experience complex nuclear synthesis processes in their cores. These processes produce heavy elements that are transported to the outer stellar atmosphere, increasing the surface metallicity and consequently reducing the [C/Fe] ratio \citep{abate2016plausible}. However, these stars exhibit higher [C/Fe] during their early evolutionary stages and are therefore still classified as CEMP stars. To distinguish these CEMP stars with surface carbon diluted by internal mixing, we have added a specific flag column in the final published candidate catalog to identify such objects.

Figure~\ref{fig:CEMP_proportion} depicts the overall trend of the CEMP star fraction as a function of metallicity. It is evident that in the low-metallicity regime, the occurrence rate of CEMP stars increases significantly and continues to rise with further decreasing [Fe/H]. This statistical feature is in good agreement with previous observational results\citep{placco2014carbon}.

\subsection{Evaluation of the CEMP Star Searching Scheme}
\label{sec:CEMPSearch_evaluation}

As the CEMP star search process presented in Section~\ref{sec:Workflow} relies on the model’s estimation of fundamental stellar parameters and chemical abundances, and the CEMP classification criteria are directly given by Eq.~\eqref{eq:CEMP_define}, the accuracy of the model-predicted $T_\texttt{eff}$, $\log~g$, [Fe/H], and [C/H] directly determines the reliability of the final CEMP identification. The estimation performance of these physical parameters is shown in Figure~\ref{fig:parameterEstimationResults}.

\begin{figure}[ht]
    \centering

    \subfloat[$T_\texttt{eff}$]{%
        \includegraphics[width=0.46\linewidth]{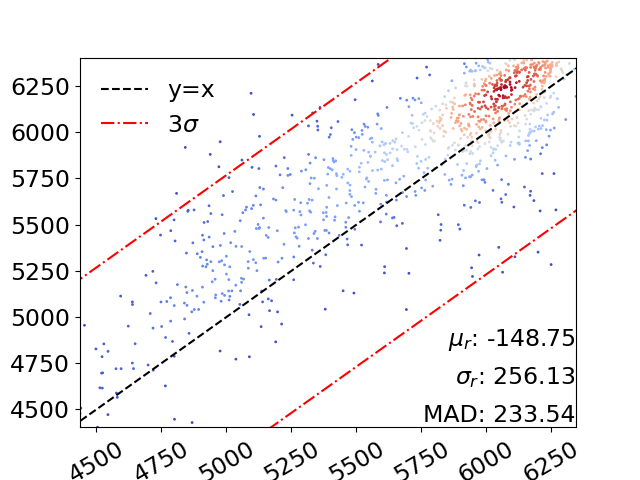}%
        \label{fig:gaia_compare_Teff}%
    }\hspace{0.01cm}%
    \subfloat[{[}Fe/H{]}]{%
        \includegraphics[width=0.46\linewidth]{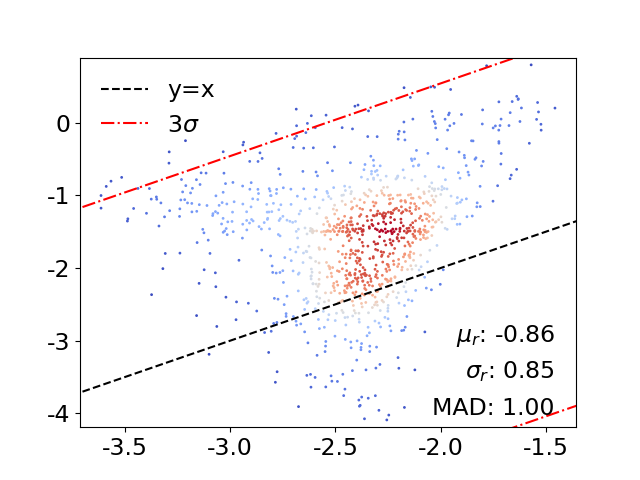}%
        \label{fig:gaia_compare_FeH}%
    }
    
    \caption{Comparison of the physical parameters of the CEMP candidates identified in this work with the corresponding parameters from Gaia DR3. The x-axis represents the parameter estimates obtained in this work, while the y-axis shows those derived from Gaia DR3. The color intensity indicates the data point density, with brighter colors corresponding to higher densities. The black dashed line denotes the line of theoretical consistency ($y = x$), and the red dashed line represents the $3\sigma$ deviation reference. The $\mu_r$ and $\sigma_r$ correspond to the mean and standard deviation of the differences between the two catalogs, respectively, while MAD denotes the mean absolute difference between the two catalogs.}
    \label{fig:gaia_compare}
\end{figure}

Building upon this, to further verify the external consistency of the physical parameters derived from low-resolution spectra, we perform a systematic comparison between the parameters of the identified CEMP candidates and those provided by Gaia DR3, as shown in Figure~\ref{fig:gaia_compare}. The results indicate good agreement in effective temperature $T_\texttt{eff}$ and metallicity [Fe/H], with only 1.84\% and 2.05\% of sources lying beyond the $3\sigma$ range, respectively. This consistency suggests that the $T_\texttt{eff}$ and [Fe/H] estimates based on LAMOST spectra are reliable, thereby providing a solid foundation for the subsequent CEMP candidate selection.

After establishing the physical of the candidates, we further quantify the CEMP searching performance from the perspective of classification accuracy. In this study, the spectra are categorized into three classes: Carbon-Enhanced Metal-Poor (CEMP) stars, carbon-normal metal-poor stars, and non-metal-poor stars, corresponding to the three-class labels ${1,2,0}$. Accordingly, we adopt multi-class evaluation metrics, including macro-averaged AUC, macro-averaged F1 score, G-mean, and Matthews correlation coefficient (MCC), to comprehensively assess the model’s discriminative capability under severe class imbalance. Further discussion of these metrics can be found in \citet{MNRAS:Zeng:2020}.

On the test set, the model achieved a macro-averaged AUC of 0.8613, a macro-averaged F1 score of 0.7781, a G-mean of 0.8526, and an MCC of 0.8578 for the three-class classification. These results indicate that, even though CEMP stars constitute a very small fraction of the overall sample, the model maintains strong discriminative capability within the three-class framework. The high G-mean demonstrates that the model can effectively identify CEMP stars corresponding to weak signals despite the highly imbalanced class structure. Meanwhile, the macro-averaged AUC close to 0.86 and MCC of 0.8578 indicate stable and consistent performance in distinguishing CEMP, Carbon-Normal Metal Poor stras, and non-metal-poor stars. Combined with a macro-averaged F1 score of 0.7781, it is evident that the model achieves a good balance between precision and recall across multiple classes. This means it can not only select a substantial fraction of CEMP stars from the abundant Carbon-Normal Metal Poor and non-metal-poor spectra, but also maintain high accuracy among the candidates identified as CEMP stars. Looking forward, as the number of confirmed CEMP stars continues to grow, expanding the training set and re-optimizing the model is expected to further enhance CEMP star identification performance.

\subsection{Classification Confidence Assessment for CEMP Star Candidates}
\label{sec:confidence_assessment}

As illustrated in Figure~\ref{fig:parameterEstimationResults}, the stellar parameters predicted by our model ($T_{\text{eff}}$, $\log~g$, [Fe/H], and [C/H]) are subject to regression uncertainties. Such uncertainties may propagate into the final CEMP classification, particularly for candidates located near the decision boundary (e.g., [C/Fe] $\sim 0.7$). To quantitatively assess this effect, we provide a probability-based confidence evaluation for all CEMP candidates.

In this work, a Monte Carlo simulation approach is adopted to estimate the classification confidence of each CEMP candidate. For each object, 50,000 random realizations of stellar parameters are generated based on the predicted values ($T_{\text{eff}}$, $\log~g$, [Fe/H], and [C/H]) and their corresponding regression errors. For each realization, the classification is recomputed following exactly the same criterion as defined in Eq.~\eqref{eq:CEMP_define}. The fraction of realizations classified as CEMP stars is then defined as the Classification Confidence of that candidate. A confidence value closer to 1 indicates that the CEMP classification is less sensitive to parameter uncertainties.

\begin{figure}[htb]
    \centering
    \setlength{\abovecaptionskip}{1pt} 
    \setlength{\belowcaptionskip}{6pt} 

    \includegraphics[width=\columnwidth]{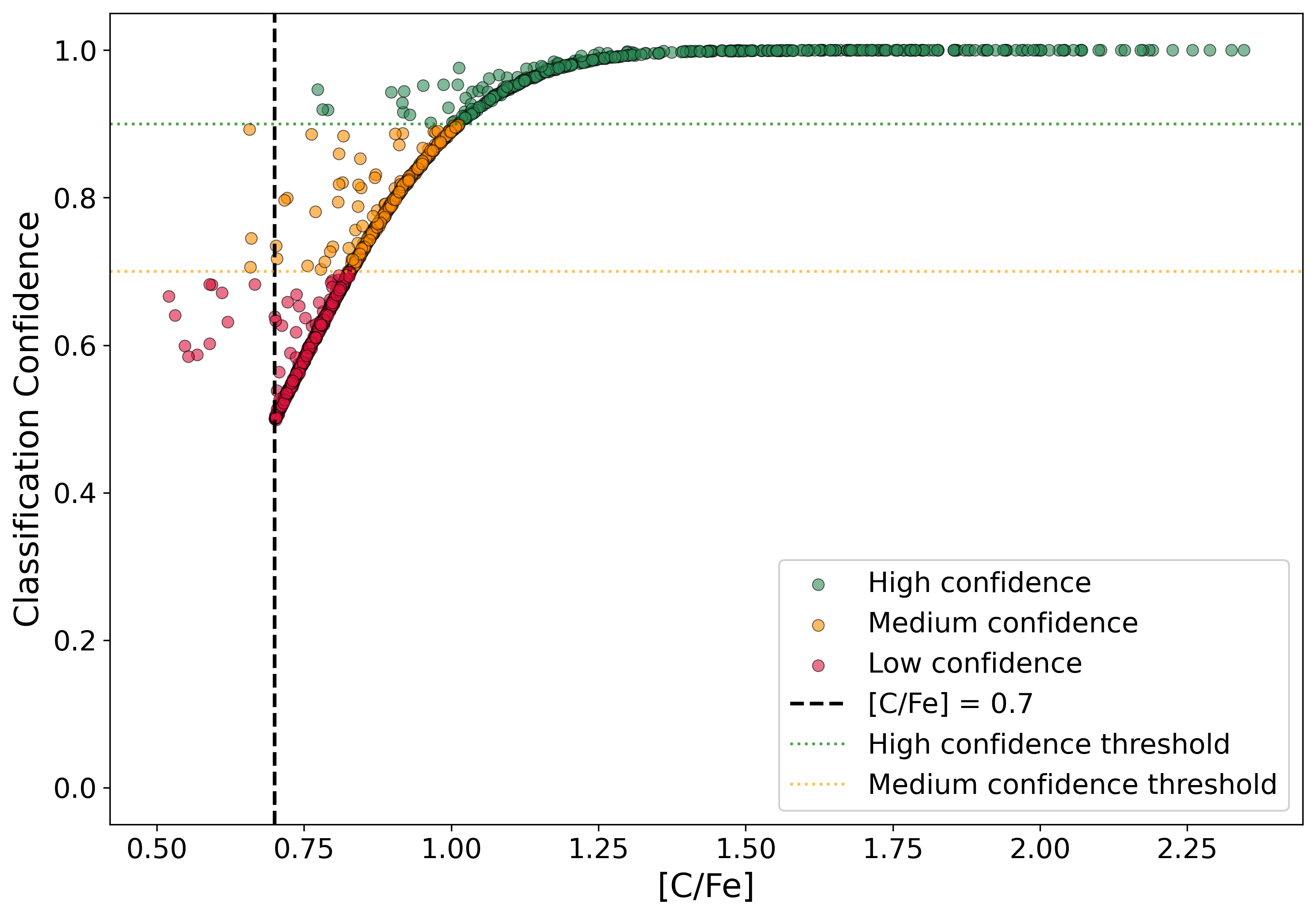}
    
    \caption{Distribution of [C/Fe] versus classification confidence for CEMP candidates. Different colors indicate high, medium, and low confidence levels.}
    \label{fig:CFe_Confidence}
\end{figure}

Among the total of 1,408 CEMP candidates, 432 objects (30.7\%) are evaluated as high-confidence candidates (confidence $\geq 0.9$), indicating that their classifications are robust against parameter uncertainties. A further 455 objects (32.3\%) are categorized as medium-confidence candidates ($0.7 \leq$ confidence $< 0.9$), suggesting moderate sensitivity to parameter errors. The remaining 521 objects (37.0\%) are classified as low-confidence candidates (confidence $< 0.7$), which are predominantly located near the boundary region of [C/Fe] $\sim 0.7$ (see Figure~\ref{fig:CFe_Confidence}).

Given the scientific importance of extremely metal-poor (EMP) stars, we further examine the confidence distribution within this subsample. Among the 104 EMP candidates identified in our search, 98 objects (94.2\%) are high-confidence candidates, 1 object (1.0\%) is a medium-confidence candidate, and the remaining 5 objects (4.8\%) are low-confidence candidates. This distribution indicates that the classification of the vast majority of EMP candidates is largely insensitive to uncertainties in the underlying stellar parameters, making them reliable targets for follow-up studies of early chemical enrichment.

Based on the above confidence assessment, the classification confidence is provided for each CEMP candidate in the final released catalog.

\subsection{CEMP Searching Performance Evaluation Based on the CH G-band}
\label{sec:CEMPSearch_EGP}

The data-driven model proposed in this work is designed for the automatic identification of CEMP stars, and its validity can be assessed through comparison with traditional spectral diagnostic indices. Studies have shown that the G1 band line index is highly responsive to carbon enhancement \citep{LiHaiNing2018ApJS}, making the flux in this band an important reference for evaluating the reliability of the model’s predictions. Meanwhile, the EGP index defined in \citet{Placco2011AJ} effectively distinguishes CEMP candidates in low-resolution spectra by comparing the relative strength of CH absorption to the $g$ band, while also mitigating contamination from the strong $H_\gamma$ line to some extent. The detailed calculation of the EGP index is given in Eq.~\eqref{eq:EGP}.
\begin{equation}
\label{eq:EGP}
\begin{aligned} 
\mathrm{EGP}=-2.5 \log \left(\frac{\int_{4200}^{4400} I_{\lambda} d \lambda}{\int_{4425}^{4520} I_{\lambda} d \lambda}\right)
\end{aligned} 
\end{equation}

\begin{figure}[ht]
    \centering
    \includegraphics[width=0.8\linewidth]{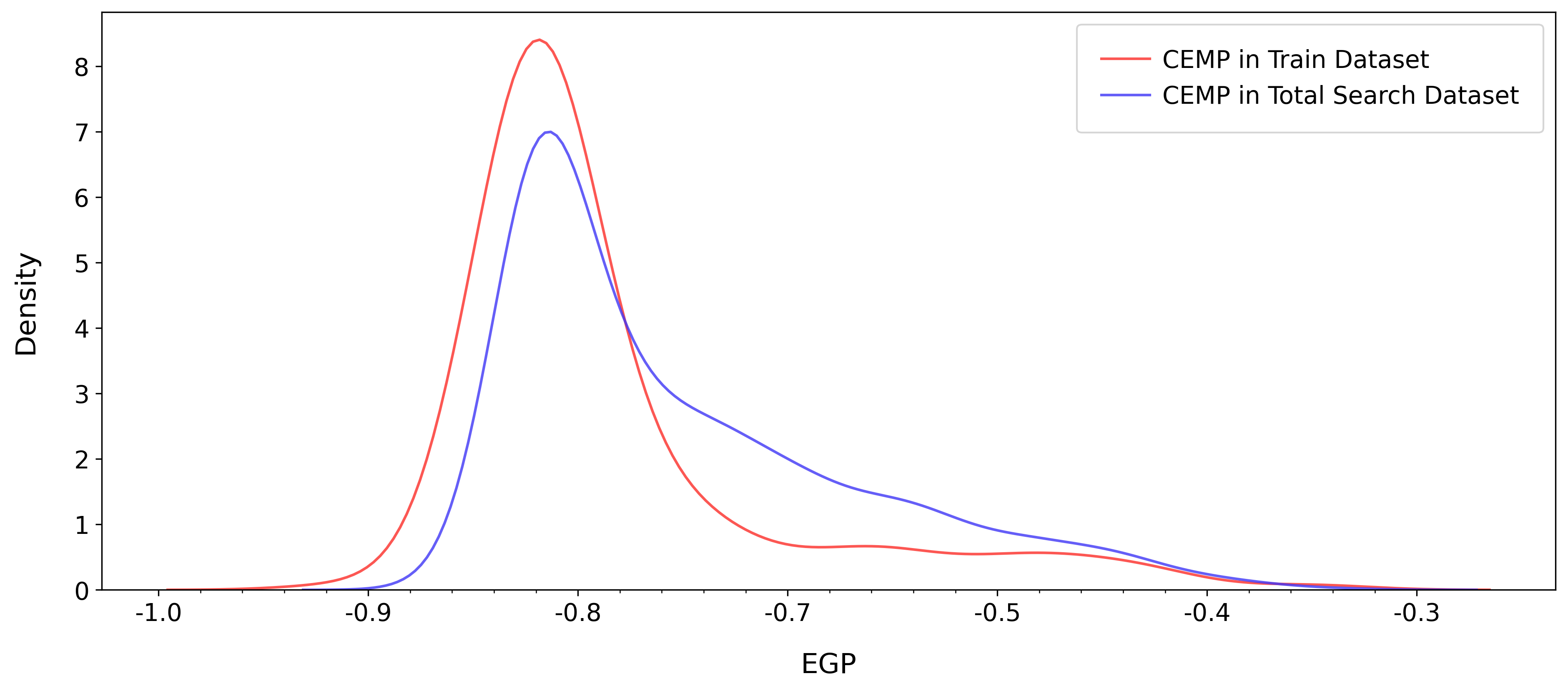}
    \caption{Distribution of EGP indices for CEMP stars, where blue denotes all CEMP stars identified in the search and red denotes CEMP stars in the training set. The horizontal axis represents the EGP value, and the vertical axis represents the density of the distribution.}
    \label{fig:EGP_rate}
\end{figure}

Figure~\ref{fig:EGP_rate} presents a comparison of the CEMP star distributions in the EGP index space, where the red curve represents the distribution of CEMP stars in the training set, and the blue curve represents the distribution of the CEMP star candidates identified in this work. The CEMP stars in the training set were carefully selected from multiple literature sources and verified through medium- or high-resolution spectroscopic observations. Both datasets exhibit a peak in sample occurrence probability within $-0.84 \leq \texttt{EGP} \leq -0.76$, and the occurrence probability of CEMP stars gradually decreases as the distance from the peak increases for $\texttt{EGP} > -0.76$ and $\texttt{EGP} < -0.84$. Therefore, the EGP occurrence ranges and distribution characteristics of the two datasets are consistent to a certain extent. This consistency indicates the reliability of the CEMP star candidates identified in this work.

\begin{figure*}[ht]
    \centering

    \subfloat[On Samples from APOGEE]{%
        \includegraphics[width=0.32\linewidth]{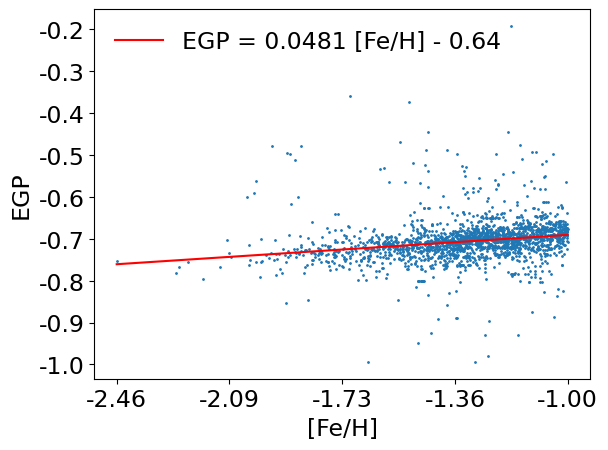}%
        \label{fig:EGP_FeH_Apogee}%
    }\hspace{0.01cm}%
    \subfloat[On Samples from VMP]{%
        \includegraphics[width=0.32\linewidth]{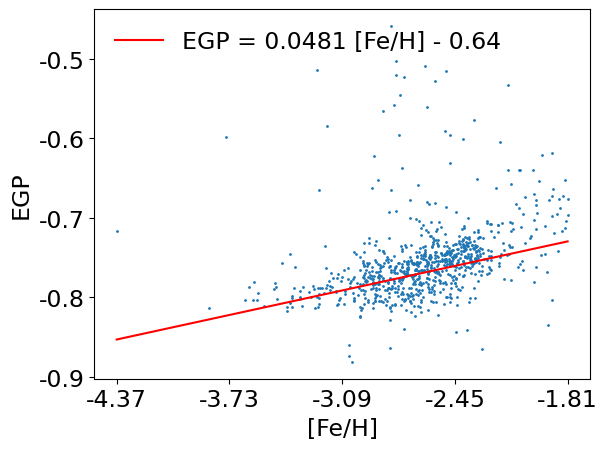}%
        \label{fig:EGP_FeH_VMP}%
    }\hspace{0.01cm}%
    \subfloat[On Samples of This Work]{%
        \includegraphics[width=0.32\linewidth]{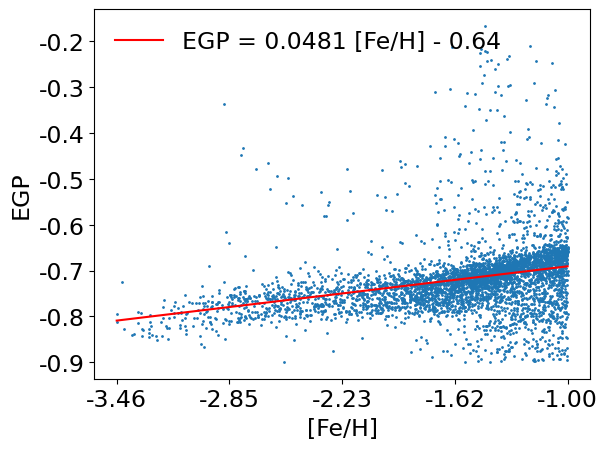}%
        \label{fig:EGP_FeH_ThisWork}%
    }

    \vspace{0.2cm}

    \subfloat[On Samples from APOGEE]{%
        \includegraphics[width=0.32\linewidth]{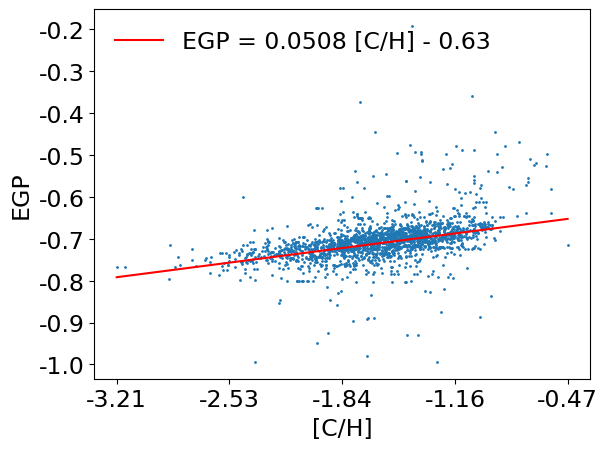}%
        \label{fig:EGP_CH_Apogee}%
    }\hspace{0.01cm}%
    \subfloat[On Samples from VMP]{%
        \includegraphics[width=0.32\linewidth]{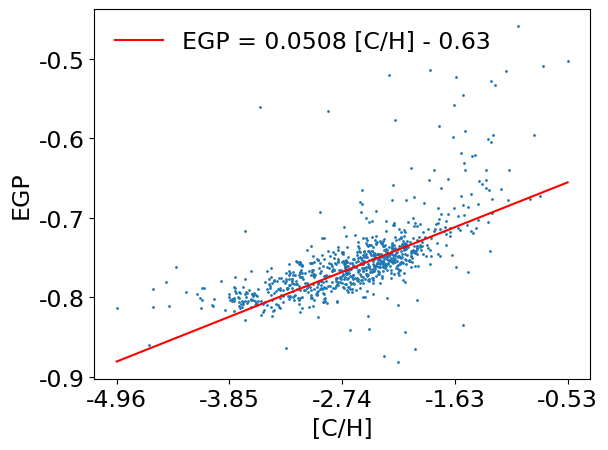}%
        \label{fig:EGP_CH_VMP}%
    }\hspace{0.01cm}%
    \subfloat[On Samples of This Work]{%
        \includegraphics[width=0.32\linewidth]{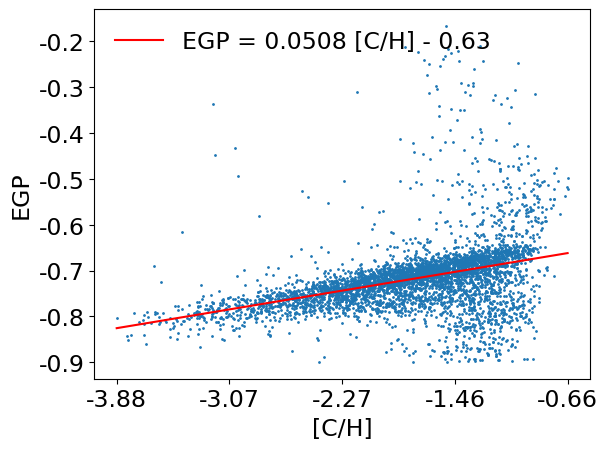}%
        \label{fig:EGP_CH_ThisWork}%
    }

    \caption{Dependence of the EGP index on [Fe/H] and [C/H] for the case of $4300~\mathrm{K} \leq T_\texttt{eff} \leq 5000~\mathrm{K}$ and $\log~g \leq 2.5$. The lines in panels (a), (b), and (c) are estimated using the samples in panel (c); similarly, the lines in panels (d), (e), and (f) are estimated in the same manner.}
    \label{fig:Dependencies_EGP_FeHCH}
\end{figure*}

Furthermore, Figure~\ref{fig:Dependencies_EGP_FeHCH} examines the dependencies of the EGP index on [Fe/H] and [C/H], respectively. The results indicate that the EGP index exhibits a good linear relationship with both [C/H] and [Fe/H] in the ranges of $4300~\text{K} \leq T_\texttt{eff} \leq 5000~\text{K}$ with $\log~g \leq 2.5$, and $T_\texttt{eff} > 5000~\text{K}$ with $\log~g \leq 4.5$.

\section{Comparing the Parameter Estimation Results with GALAH}
\label{sec:ParameterEvaluation}

The Third Data Release of the Galactic Archaeology with HERMES (GALAH DR3) \citep{buder2021galah+} provides a high-resolution spectroscopic reference sample comprising approximately 588,571 stars. This survey covers a wide range of stellar evolutionary stages, from dwarfs to supergiants, and spans a broad metallicity range from extremely metal-poor to metal-rich stars. The GALAH spectra have a resolving power of up to 28,000, enabling precise measurements of the abundances of up to 30 chemical elements. Stellar atmospheric parameters are initially estimated through global fitting based on the AMBRE spectral grid and are subsequently refined by linearly combining the parameters of the ten closest synthetic spectra in parameter space.

\begin{figure}[ht]
    \centering

    \begin{minipage}[t]{0.24\linewidth}
        \centering
        \includegraphics[width=\linewidth]{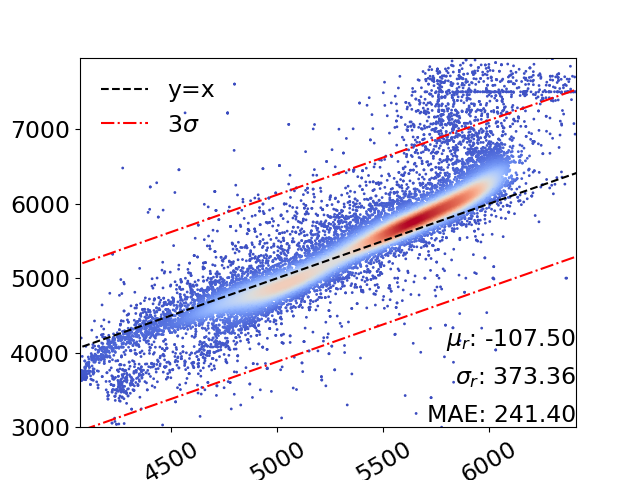}
        \small (a) $T_\texttt{eff}$
    \end{minipage}\hspace{0.01cm}
    \begin{minipage}[t]{0.24\linewidth}
        \centering
        \includegraphics[width=\linewidth]{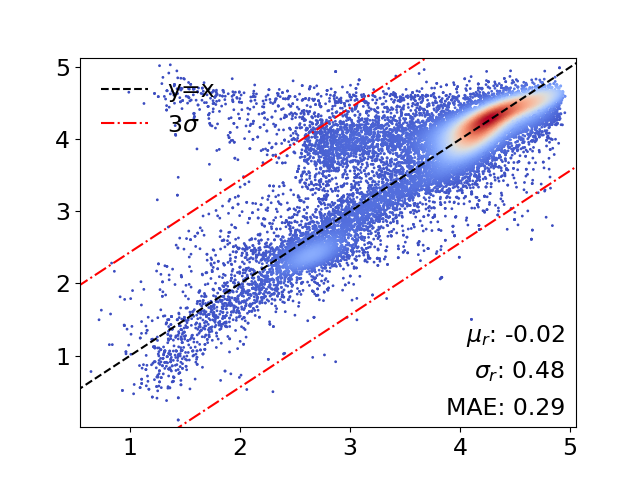}
        \small (b) $\log~g$
    \end{minipage}\hspace{0.01cm}
    \begin{minipage}[t]{0.24\linewidth}
        \centering
        \includegraphics[width=\linewidth]{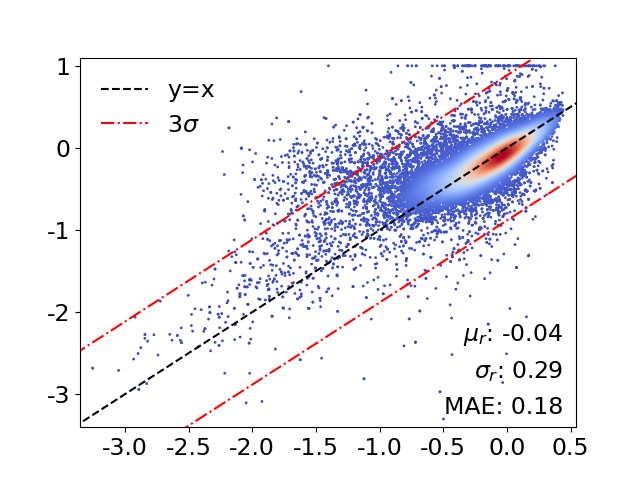}
        \small (c) [Fe/H]
    \end{minipage}\hspace{0.01cm}
    \begin{minipage}[t]{0.24\linewidth}
        \centering
        \includegraphics[width=\linewidth]{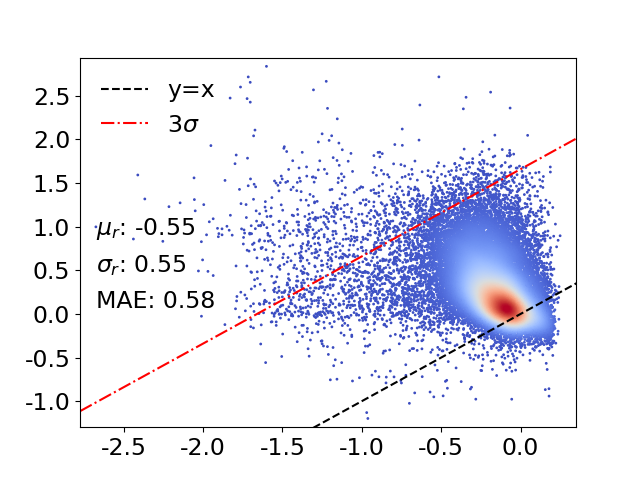}
        \small (d) [C/H]
    \end{minipage}

    \caption{Comparison between this work and GALAH DR3. The x-axis represents the parameter estimates from this work, while the y-axis represents those from GALAH DR3. The color intensity corresponds to the sample density, with brighter colors indicating higher densities. The black dashed line denotes the line of theoretical consistency, whereas the red dashed line indicates the $3\sigma$ reference line. The $\mu_r$ and $\sigma_r$ represent the mean and standard deviation of the differences between the two catalogs, respectively. MAE denotes the mean absolute difference between the two catalogs.}
    \label{fig:galah_compare}
\end{figure}

According to the CEMP star classification criterion defined in Eq.~\eqref{eq:CEMP_define}, the identification of CEMP stars critically depends on the accurate estimation of four key stellar parameters: $T_\texttt{eff}$, $\log~g$, [Fe/H], and [C/H]. Therefore, consistency with parameter estimates derived from high-resolution spectroscopic surveys serves as an important benchmark for assessing the reliability of the proposed method. In Section~\ref{sec:parameterEstimation_evaluation}, the parameter estimates obtained in this work exhibit a high level of agreement with several authoritative high-resolution stellar catalogs, including APOGEE DR17 \citep{ApJS:Abdurro:2022}, the LAMOST–Subaru catalog \citep{Aoki2022ApJ, LiHaiNing2022ApJ}, the SAGA database \citep{PASJ:Suda:2008, MNRAS:Suda:2011, MNRAS:Suda:2013, PASJ:Suda:2017}, and the VMP catalog \citep{yuan2020dynamical}. Building upon the analysis in Section~\ref{sec:CEMPSearch_evaluation}, we further compare our parameter estimation results with those from GALAH DR3\citep{buder2021galah+}, thereby providing an independent and indirect validation of the robustness of our CEMP star search results. A detailed comparison between GALAH and our results is presented in Figure~\ref{fig:galah_compare}. To quantitatively assess the consistency with GALAH DR3, we calculated the median bias and median absolute deviation (MAD) for each parameter, with the results summarized in Table~\ref{tab:galah_bias_stats}.

In the comparison of effective temperature ($T_\texttt{eff}$), we observe a divergence branch in our results relative to GALAH when the GALAH parameter distribution lies within the interval of [6773\,K, 8000\,K] (see Figure~\ref{fig:galah_compare}(a)). This behavior primarily originates from differences in the coverage of the training samples: the training data used in this work are mainly concentrated in the range of [3609\,K, 6907\,K], and for stars with higher effective temperatures beyond this interval, the estimated temperatures may be subject to a certain degree of underestimation. As shown in Table~\ref{tab:galah_bias_stats}, the overall median bias for $T_{\text{eff}}$ is -72.7 K, with a median absolute deviation of 144.3 K, which primarily stems from systematic underestimation at the high-temperature end. It is worth noting that only about 0.14\% of the samples exhibit significant temperature discrepancies, and these samples are almost exclusively located at the high-temperature end. No CEMP stars are identified in this temperature regime, which is consistent with physical expectations. As $T_\texttt{eff}$ increases, carbon-related spectral features become significantly weaker, in agreement with the findings of Witten et al.\ \citep{witten2022information}, who showed that the precision of [C/H] measurements can reach 0.25 dex only when $T_\texttt{eff}<6000~\mathrm{K}$. Consequently, the discrepancies between this work and GALAH in the high-temperature regime do not have a substantive impact on the practical search for CEMP stars.

Regarding the estimation of surface gravity ($\log~g$), our results show a high overall consistency with those from GALAH, with only about 0.12\% of stars exhibiting a slight underestimation near $\log~g \approx 5.0$ dex. Quantitative statistics (Table~\ref{tab:galah_bias_stats}) reveal that the median bias for $\log~g$ is only 0.055 dex, with a median absolute deviation of 0.166 dex, further confirming the agreement of the estimates. No CEMP stars are detected in the divergence region, which is likewise expected. In our reference set, the maximum $\log~g$ value of CEMP stars is approximately 4.5 dex; therefore, for potential CEMP stars with higher surface gravities, the model may be subject to a certain risk of misclassification.

In terms of metallicity ([Fe/H]) estimation, our results show a high level of consistency with GALAH in the overall distribution, particularly in the extremely metal-poor regime with [Fe/H] $< -2$. This agreement indicates that the very metal-poor (VMP) candidates selected in this work are of high credibility. Quantitative analysis further indicates (Table~\ref{tab:galah_bias_stats}) that the overall median bias for [Fe/H] is only 0.011 dex, with a median absolute deviation of 0.105 dex, demonstrating no significant systematic bias compared to GALAH DR3. Given that CEMP stars are predominantly found among the VMP population, the high accuracy in VMP identification further supports the reliability of the CEMP stars identified in our search.

For carbon abundance ([C/H]), a certain systematic offset is observed between our results and those from GALAH. Quantitative analysis (Table~\ref{tab:galah_bias_stats}) reveals a median bias of -0.433 dex and a median absolute deviation of 0.328 dex for [C/H], indicating a systematic underestimation. The systematic offset in [C/H] directly affects the derived [C/Fe] = [C/H] – [Fe/H], the key quantity in the CEMP classification criterion (Eq.~\eqref{eq:CEMP_define}). In principle, an underestimation of [C/H] by $\sim$0.43 dex would lead to a corresponding underestimation of [C/Fe]. This systematic lowering of [C/Fe] could, in theory, cause stars with intrinsic [C/Fe] just above the threshold (e.g., [C/Fe] $\approx$ 0.7) to be misclassified as non‑CEMP. Theoretically, such a bias could affect stars near the classification threshold at any metallicity. In practice, however, CEMP stars in the very metal-poor regime ([Fe/H] $< -2$) typically exhibit [C/Fe] values substantially above the threshold, making them less susceptible to this underestimation—though the possibility of missing borderline stars cannot be entirely ruled out. The CEMP classification also depends on $\log(L/L_\odot)$, derived from $T_{\mathrm{eff}}$ and $\log~g$. As shown in the preceding comparison with GALAH DR3, our estimates of $T_{\mathrm{eff}}$, $\log~g$, and [Fe/H] exhibit excellent agreement with GALAH within the parameter space where CEMP stars are typically found (median biases of $-$72.7~K, 0.055~dex, and 0.011~dex, respectively). Hence, the potential impact on CEMP selection is dominated by the [C/H] offset. To evaluate whether this theoretical concern translates into a practical effect, we examined the cross‑matched sample with GALAH, focusing on stars with large [C/H] deviations or those near the CEMP classification boundary. Remarkably, neither subset contains any CEMP stars identified by either our method or GALAH. Furthermore, no CEMP stars are found in the region where the [C/H] differences are substantial. Therefore, despite the systematic offset, it does not affect the identification of CEMP candidates within our sample. It is worth noting, however, that similar systematic differences are also present in the comparison between APOGEE DR17 and GALAH. Such offsets are more likely attributable to differences in carbon abundance calibration standards and modeling strategies adopted by different surveys, rather than to intrinsic deficiencies in the methodology proposed in this work.

In summary, this work demonstrates a high degree of consistency with the GALAH results within the parameter ranges of effective temperature $T_{\texttt{eff}} \in [3609.00, 6907.00]~\mathrm{K}$, surface gravity $\log~g \in [-0.14, 5.08]~\mathrm{dex}$, metallicity $\mathrm{[Fe/H]} \in [-4.37, 0.50]~\mathrm{dex}$, and carbon abundance $\mathrm{[C/H]} \in [-5.01, 0.90]~\mathrm{dex}$. In particular, the accuracy of parameter estimation for very metal-poor (VMP) stars further supports the reliability of the Carbon-Enhanced Metal-Poor (CEMP) stars identified in this study.

\begin{table}[htbp]
    \centering
    \caption{Systematic bias statistics in comparison with GALAH DR3}
    \label{tab:galah_bias_stats}
    
    \begin{tabular}{lcc}
    \hline\noalign{\smallskip}
    Parameter & Median Bias & MAD \\
    \noalign{\smallskip}\hline\noalign{\smallskip}
    
    $T_{\mathrm{eff}}$ & -72.7 K & 144.3 K \\
    $\log~g$ & 0.055 dex & 0.166 dex \\
    $[\mathrm{Fe/H}]$ & 0.011 dex & 0.105 dex \\
    $[\mathrm{C/H}]$ & -0.433 dex & 0.328 dex \\
    
    \noalign{\smallskip}\hline
    \end{tabular}
    
    \smallskip
    \parbox{0.4\columnwidth}{ 
    \footnotesize
    \textit{Note.} Median Bias is the median difference between our estimated values and the GALAH DR3 reference values, with positive values indicating overestimation and negative values indicating underestimation. MAD is the median absolute deviation, measuring dispersion.}
    
\end{table}

\section{Conclusion}
\label{sec:conclusion}

This work focuses on the search for Carbon-Enhanced Metal-Poor (CEMP) stars under low-resolution stellar spectroscopic conditions and constructs a data-driven analysis framework centered on a parallel dual-view deep network combined with an ensemble learning strategy. The proposed method enables stable inversion of key stellar physical parameters, including $T_\texttt{eff}$, $\log~g$, [Fe/H], and [C/H], directly from observed spectra, and subsequently facilitates efficient identification of CEMP star candidates. When applied to the low-resolution spectral sample from LAMOST DR11, the framework identifies a total of 1,408 CEMP star candidates, comprising 1,302 VMP stars and 104 EMP star candidates. In addition to the candidate catalog, this study also provides the estimated $T_\texttt{eff}$, $\log~g$, [Fe/H], and [C/H] parameters for each target, thereby offering directly usable reference data for subsequent statistical analyses and target selection for high-resolution spectroscopic follow-up observations.

It should be noted that, owing to the intrinsic information limitations of low-resolution LAMOST spectra, the framework proposed in this work is not yet able to reliably distinguish between different physical sub-classes of CEMP stars, such as CEMP-s and CEMP-no. This limitation primarily arises from the lack of diagnostic spectral features of key neutron-capture elements (e.g., Ba, Sr, and Eu) at low spectral resolution, rather than from deficiencies in the model architecture or learning strategy itself. Consequently, the present results are best interpreted as an efficient candidate selection tool for subsequent high-resolution follow-up observations, rather than as a definitive classification of CEMP sub-classes.

Among the 1,408 CEMP star candidates identified in this study, particular attention should be paid to the 104 extremely metal-poor (EMP) star candidates, which represent some of the most promising targets for probing early chemical enrichment and the nature of the first generations of stars. These objects are especially valuable for investigations of primordial nucleosynthesis processes and the early formation history of the Galaxy. For these most compelling candidates, systematic high-resolution spectroscopic follow-up is strongly encouraged. Such observations would enable robust constraints on neutron-capture elements (e.g., Ba, Sr, and Eu), together with detailed characterization of carbon-sensitive molecular features (e.g., CH and C$_2$ bands), thereby allowing reliable discrimination between CEMP-s, CEMP-no, and related sub-classes. When combined with the stellar parameter estimates and candidate catalog provided in this work, these follow-up strategies can substantially enhance the scientific return while maintaining observational efficiency. The computed catalog is available via \href{https://doi.org/10.12149/101799}{doi:10.12149/101799}.

\noindent\textbf{Acknowledgements} ~~~~This work is supported by  the National Natural Science Foundation of China (Grants No. 12373108).

\bibliographystyle{raa}
\bibliography{bibtex}     

\label{lastpage}

\end{document}